\documentclass[12pt]{article}
\usepackage[english]{babel}
\usepackage[pdftex]{color}
\usepackage{amsmath}
\usepackage{graphicx}
\usepackage{hyperref}
\usepackage{amsmath}
\usepackage{amsfonts}
\usepackage{amssymb}

\begin{document}
\date{}

\begin{center}
{\Large\textbf{{}General Cubic Interacting Vertex\\ for Massless
Integer Higher Spin Fields}}
\vspace{18mm}

{\large I.L. Buchbinder$^{(a,b)}\footnote{E-mail:
joseph@tspu.edu.ru}$
,\;
A.A. Reshetnyak$^{(a,b)}\footnote{E-mail:
reshet@tspu.edu.ru}$}

\vspace{8mm}

\noindent  ${{}^{(a)}} ${\em
Center of Theoretical Physics, \\
Tomsk State Pedagogical University,\\
Kievskaya St.\ 60, 634061 Tomsk, Russia}

\noindent  ${{}^{(b)}} ${\em
National Research Tomsk State  University,\\
Lenin Av.\ 36, 634050 Tomsk, Russia}

\vspace{20mm}

\begin{abstract}
We consider a massless higher spin field theory within the BRST
approach and construct a general off-shell cubic vertex
corresponding to irreducible higher spin fields of helicities $s_1,
s_2, s_3$. Unlike the previous works on cubic vertices, which do not
take into account of the trace constraints, we use the complete BRST
operator, including the trace constraints that describe an
irreducible representation with definite integer helicity. As a
result, we generalize the cubic vertex found in \cite{BRST-BV3} and
calculate the new contributions to the vertex, which contain
additional terms with a smaller number space-time derivatives of the
fields as well as the terms without derivatives.
\end{abstract}

\end{center}
\thispagestyle{empty}

\section{Introduction}

Study of the various aspects of the higher spin fields theory is one
of the topical trends in modern high-energy theoretical and
mathematical physics (for a review, see, e.g. \cite{reviews},
\cite{revvas},  \cite{revBCIV},  \cite{reviews3}, \cite{rev_Bekaert}, \cite{revDS},
\cite{reviewsV}
and the references therein). The current progress in higher spin field
theory is closely related to a description of interactions between
these fields.  It is expected that taking account of interacting
higher spin fields will open up the new perspectives of going beyond
the Standard Model and contribute to the formation of new approaches
to unifying the all fundamental interactions, including quantum
gravity.

In this letter, we focus on some new aspects involving a cubic
vertex for massless integer higher spin fields. The structure of
cubic vertices has been investigated using different approaches in
numerous works (see, e.g., the recent papers \cite{Manvelyan},
\cite{Manvelyan1}, \cite{Joung}, \cite{frame-like1},
\cite{Metsaev0712}, \cite{BRST-BV3},  \cite{frame-like2}, \cite{BKTW}
and the references therein); however, in these papers the vertex was
constructed using fields  subject the algebraic constraints which are not derived from the action principle.  In our opinion, almost all of
the results known to date on the structure of cubic vertices are
contained in a concise form in the work \cite{BRST-BV3}, where the
cubic vertex was deduced using the BRST approach, and the results
are consistent with the light-cone consideration \cite{Metsaev0512}.
\footnote{It is worth pointing out
the interacting higher spin field models (see \cite{Skvortsov} and
the references therein) containing the cubic vertices, which possess
the remarkable properties in quantum domain.}.

In general, a central object of BRST approach, the BRST operator is
constructed using operator constraints which form a first-class
gauge algebra. In the case under consideration, the corresponding
constraints should include a dynamical on-shell condition $l_0$ and
constraints $l_1, \, l_{11}$, responsible for divergences and
traces, respectively. The explicit forms of $l_0,\, l_1,\, l_{11}$
are given by (\ref{irrepint}). However, in many cases (see, e.g.,
\cite{reviews3}, \cite{BRST-BV3} and references therein) the BRST
charge is constructed, for the simplicity of calculations, without
any trace constraints, with these constraints being imposed
afterwards by hands. This consideration is certainly correct;
however, the actual Lagrangian description of irreducible fields is
achieved only after imposing the subsidiary conditions which are not
derived from the Lagrangian. As a result, we face the problem of
constructing a cubic vertex on the base of  BRST operator including
all the constraints. It is precisely this problem that is solved in
this letter. One can expect the final cubic vertex to contain, as
distinct from \cite{BRST-BV3}, some new terms containing the trace
constraints.

The BRST approach to a Lagrangian description of various free higher
spin field models in Minkowski and AdS spaces has been developed in
numerous works (e.g., see \cite{PT}, \cite{BPT}, \cite{BR} and the
review \cite{reviews3}). The development of the BRST approach for
calculating the cubic vertex was initiated by paper \cite{BFPT}.
Now, we intend to present a complete solution of this problem and
describe a general structure of the vertex, while taking into
account all the constraints required to formulate an irreducible
representation of the Poincare group within the BRST approach.

The paper is organized as follows. Section~\ref{cBRSTBFV} presents
the basics of a BRST Lagrangian construction for free massless
higher spin field, with all the constraints $l_0,\, l_1,\, l_{11}$
taken into account. In Section~\ref{BRSTinter}, we deduce a system
of equations for a cubic (linear) deformation in fields of the free
action (free gauge transformations). A solution for the deformed
cubic vertices and gauge transformation is given in
Section~\ref{intBRSTBFV}. The main result of the work is that the
cubic vertex and deformed gauge transformations include both the
constraint $l_{1}$ and the constraint $l_{11}$. Conclusion gives a
final summary and comments.

We use the standard definition $\eta_{\mu\nu} = diag (+, -,...,-)$
for a metric tensor with Lorentz indices $\mu, \nu =
0,1,...,d-1$ and the respective notation $\epsilon(F)$, $gh(F)$,
$[F,\,G\}$, $[x]$ for the values of Grassmann parity and
ghost number of a homogeneous quantity $F$, as well as
the supercommutator and the integer part of a real-valued $x$.

\section{BRST Lagrangian formulation for free higher spin  fields}

\label{cBRSTBFV}

In this section, we briefly present the basic results  of the BRST
approach to free massless higher integer spin field theory. All
these results will be used in the next section to describe a general
structure of cubic interacting vertex.

As known, the unitary massless irreducible representations of
Poincare $ISO(1,d-1)$ group with integer helicities $s$ can be
realized using the real-valued totally symmetric tensor fields
$\phi_{\mu_1...\mu_s}(x)\equiv \phi_{\mu(s)}$ under the following
conditions
\begin{eqnarray}\label{irrepint}
    &&  \big(\partial^\nu\partial_\nu,\, \partial^{\mu_1},\, \eta^{\mu_1\mu_2}\big)\phi_{\mu(s)}  = (0,0,0)  \  \ \ \   \Longleftrightarrow  \  \\
     &&       \big(l_0,\, l_1,\, l_{11}, g_0 -d/2\big)|\phi\rangle  = (0,0,0,s)|\phi\rangle. \nonumber
\end{eqnarray}
Here the basic vectors $|\phi\rangle$ and the operators $l_0,\,
l_1,\, l_{11}, g_0 -d/2$ are defined in the Fock space $\mathcal{H}$
with the bosinic oscillators $a_\mu, a^+_\nu$, ($[a_\mu, a^+_\nu]= - \eta_{\mu\nu}$)  in the form
\begin{eqnarray}\label{FVoper}
&&   |\phi\rangle  =  \sum_{s\geq 0}\frac{\imath^s}{s!}\phi^{\mu(s)}\prod_{i=1}^s a^+_{\mu_i}|0\rangle, \\
&&   \big(l_0,\, l_1,\, l_{11}, g_0\big) = \big(\partial^\nu\partial_\nu ,\, - \imath a^\nu  \partial_\nu ,\, \frac{1}{2}a^\mu a_\mu ,  -\frac{1}{2}\big\{a^+_{\mu},\, a^{\mu}\big\}\big).\nonumber
\end{eqnarray}
Within the BRST approach, the free dynamics of the field with
definite helicity $s$  is described by the gauge invariant action
depending on basic field $\phi_{\mu(s)}$ and twelve  auxiliary
fields $\phi_{1\mu(s-1)},...$ of lesser than $s$ ranks. All these
fields are incorporated into the vector $|\chi\rangle_s$ and
described by the action
\begin{eqnarray}
\label{PhysStatetot} \mathcal{S}_{0|s}[\phi,\phi_1,...]=
\mathcal{S}_{0|s}[|\chi\rangle_s] = \int d\eta_0 {}_s\langle\chi|
KQ|\chi\rangle_s,
\end{eqnarray}
where $\eta_0$ be a ghost field and $K$ be an operator defining the
inner product. The action (\ref{PhysStatetot}) is invariant under
the gauge transformations
\begin{eqnarray}
\label{gauge trasnform}
\delta|\chi\rangle_s =  Q|\Lambda\rangle_s ,
\end{eqnarray}
where the gauge parameter vector $|\Lambda\rangle_s$ is defined up to the gauge
transformations
\begin{eqnarray}
\delta |\Lambda\rangle_s = Q|\Lambda^1\rangle_s
,  \ \ \delta |\Lambda^1\rangle_s =0.
\label{SunBRSTclas}
\end{eqnarray}
Here the vectors $|\Lambda\rangle_s$, $|\Lambda^1\rangle_s$ are the
vectors of zero-level and first-level gauge  parameters of the
abelian gauge transformations (\ref{SunBRSTclas}), which reflect the
fact that the theory is the first-stage reducible gauge theory. The
quantity $Q$ in (\ref{PhysStatetot}) is the  BRST operator,
constructed on the base of the constraints $l_0,\, l_1,\, l^{+}_1,\,
l_{11},\, l^{+}_{11}=\frac{1}{2}a^{+\nu}a^{+}_{\nu}$ and contains the
anticommuting ghost operators $\eta_0,\, \eta_1^+,\, \eta_1,\,
\eta_{11}^+,\, \eta_{11},\, {\cal{}P}_0,\, \mathcal{P}_{1},\,
\mathcal{P}^+_{1},$ $\mathcal{P}_{11},\, \mathcal{P}^+_{11},$
\begin{eqnarray}
\hspace{-0.5ex}&\hspace{-0.5ex}&\hspace{-0.5ex} {Q} =
\eta_0l_0+\eta_1^+l_1+l_1^{+}\eta_1+
\eta_{11}^+\widehat{L}_{11}+\widehat{L}_{11}^{+}\eta_{11} +
{\imath}\eta_1^+\eta_1{\cal{}P}_0
\label{Qctotsym}\\
\hspace{-0.5ex}&\hspace{-0.5ex}&\hspace{-0.5ex} \phantom{Q} =
\eta_0l_0+\Delta Q  + {\imath}\eta_1^+\eta_1{\cal{}P}_0  ,
\end{eqnarray}
where
\begin{eqnarray}
\hspace{-0.5ex}&\hspace{-0.5ex}&\hspace{-0.5ex}
\big(\widehat{L}_{11} ,\,\widehat{L}{}^+_{11}\big) =  \big(
L_{11}+\eta_{1} \mathcal{P}_{1} , \,
L^+_{11}+\mathcal{P}^+_{1}\eta^+_{1} \big).
\label{extconstsp2}
\end{eqnarray}
Here
\begin{eqnarray}
L_{11}=l_{11}+(b^+b+h)b,\,\, \ \  L^{+}_{11}=l^+_{11}+b^+
\end{eqnarray}
and $(\epsilon, gh) Q = (1, 1).$ The algebra of the operators $l_0$ ,$l_1$, $l^{+}_1, L_{11}, L_{11}^+,
G_0$ looks like
\begin{equation}\label{subalgebr}
[l_0, l^{(+)}_1] = 0, \ [l_1,l_1^+]=l_0 \quad \mathrm{and} \quad  [L_{11},  L_{11}^+] = G_0,\
[G_0, L_{11}^{+}] = 2L_{11}^+
\end{equation}
and their independent non-vanishing  cross-commutators are $[l_1,L_{11}^+]=-l_1^+$,  $[l_1,G_{0}]=l_1$.

The parameter $h = h(s)=-s - \frac{d-6}{2}.$
The ghost operators satisfy the following non-vanishing anticommutation relations
\begin{equation}\label{ghanticomm}
  \{\eta_0, \mathcal{P}_0\}= \imath,\ \  \ \{\eta_1, \mathcal{P}_1^+\}=\{\eta^+_1, \mathcal{P}_1\}= \{\eta_{11},   \mathcal{P}_{11}^+\}=\{\eta_{11}^+,   \mathcal{P}_{11}\}=1.
\end{equation}

The theory under consideration is characterized by the spin operator
${\sigma}$, which is defined as follows
\begin{eqnarray}
\hspace{-0.5ex}&\hspace{-0.5ex}&\hspace{-0.5ex}  {\sigma}  =   G_0+ \eta_1^+\mathcal{P}_{1}
-\eta_1\mathcal{P}_{1}^+  + 2(\eta_{11}^+\mathcal{P}_{11} -\eta_{11}\mathcal{P}_{11}^+) .
\label{extconstsp3}
\end{eqnarray}
Here $G_0=g_0 +2b^+b+h$ with $b,\, b^+$ ($[b,\, b^+]=1$) be auxiliary bosonic oscillators.
The operator ${\sigma}$ selects the vectors
with definite spin value $s$
\begin{eqnarray}
 \hspace{-0.5ex}&\hspace{-0.5ex}&\hspace{-0.5ex} {\sigma} (|\chi\rangle_s,\,
 |\Lambda\rangle_s,\, |\Lambda^1\rangle_s)  = (0,0,0),
\label{extconstsp}
\end{eqnarray}
where the Grassmann parities and the ghost numbers of the above vectors are $(0,0)$, $(1,-1),$ $(0,-2)$
respectively.

All the operators  above act  in a total Hilbert space with the inner product of the vectors depending
on all oscillators and ghosts
\begin{eqnarray}
&& \langle\chi |\psi\rangle = \int d^d x \langle0|  \chi^*\big(a,b;\eta_1, \mathcal{P}_1,\eta_{11}, \mathcal{P}_{11}\big)\psi\big(a^+,b^+;\eta^+_1, \mathcal{P}^+_1,\eta^+_{11}, \mathcal{P}^+_{11}\big)|0\rangle.
\label{scalarprod}
\end{eqnarray}

The operators $Q, {\sigma}$ are supercommuting and  Hermitian with
respect to the scalar product (\ref{scalarprod}) including the
operator $K$ (see e.g., \cite{BPT}, \cite{BR})
  \begin{align}\label{geneq}
   & Q^2 =  \eta_{11}^+\eta_{11} \sigma ,\ &&  [Q,\, \sigma\} =0; \\
   & Q^+K =  KQ ,  \ && \sigma^+K =  K\sigma ,  \label{geneq1}\\
   &    K= \sum_{n=0}^{\infty}\frac{1}{n!}(b^+)n|0\rangle\langle 0|b^n
   \prod_{i=0}^{n-1}(i+h(s))&&
   \label{geneq2}
  \end{align}
The BRST operator $Q$ is nilpotent on the subspace with zero
eigenvectors for the spin operator $\sigma$ (\ref{extconstsp}).

The field $ |\chi\rangle_s$, the zero $|\Lambda\rangle_s$ and  the
first $|\Lambda^1\rangle_s$ level gauge parameters
 labeled by the symbol $"s"$ as eigenvectors
of the spin condition in  (\ref{extconstsp}) can be written in the
form
\begin{eqnarray}
\hspace{-1em}&\hspace{-1em}&\hspace{-1em} |\chi\rangle_s  =
|\phi\rangle_s+\eta_1^+\Big(\mathcal{P}_1^+|\phi_2\rangle_{s-2}+\mathcal{P}_{11}^+|\phi_{21}\rangle_{s-3} +\eta_{11}^+\mathcal{P}_1^+\mathcal{P}_{11}^+|\phi_{22}\rangle_{s-6}\Big) \label{spinctotsym} \\ \hspace{-1em}&\hspace{-1em}&\hspace{-1em} \phantom{ |\chi^0_c\rangle_s} +\eta_{11}^+\Big(\mathcal{P}_1^+|\phi_{31}\rangle_{s-3}+\mathcal{P}_{11}^+|\phi_{32}\rangle_{s-4}\Big)+  \eta_0\Big(\mathcal{P}_1^+|\phi_1\rangle_{s-1}+\mathcal{P}_{11}^+|\phi_{11}\rangle_{s-2}  \nonumber \\ \hspace{-1em}&\hspace{-1em}&\hspace{-1em} \phantom{ |\chi^0_c\rangle_s}  +  \mathcal{P}_1^+\mathcal{P}_{11}^+\Big[ \eta^+_{1} |\phi_{12}\rangle_{s-4}+\eta^+_{11} |\phi_{13}\rangle_{s-5}\Big]\Big),\nonumber \\
\hspace{-1em}&\hspace{-1em}&\hspace{-1em} |\Lambda\rangle_s  =  \mathcal{P}_1^+  |\xi\rangle_{s-1}+\mathcal{P}_{11}^+|\xi_{1}\rangle_{s-2} +\mathcal{P}_1^+\mathcal{P}_{11}^+\Big(\eta_1^+|\xi_{11}\rangle_{s-4} \label{parctotsym}
\\
\hspace{-1em}&\hspace{-1em}&\hspace{-1em} \phantom{|\chi^1\rangle_s}
 + \eta_{11}^+|\xi_{12}\rangle_{s-5}\Big) +  \eta_0\mathcal{P}_1^+\mathcal{P}_{11}^+|\xi_{01}\rangle_{s-3} ,
  \nonumber\\
\hspace{-1em}&\hspace{-1em}&\hspace{-1em} {|\Lambda^1\rangle_s}  =
\mathcal{P}_1^+\mathcal{P}_{11}^+|\xi^{1}\rangle_{s-3}.
\end{eqnarray}
One can prove that after imposing the appropriate gauge conditions
and eliminating the auxiliary fields from equations of motion, the
theory under consideration is reduced to Fronsdal form \cite{Fronsdal} in terms of
totally symmetric double traceless tensor field $\phi_{\mu(s)}$  and
traceless gauge parameter $\xi_{\mu(s-1)}$.

Now we  pass to the construction of the interacting theory.

\section{Including the interaction: system of equations for cubic vertex}
\label{BRSTinter}

In this section we describe the general scheme of finding the cubic
interaction vertices for the theory under consideration and derive
the equations for these vertices.

Including the cubic interaction means the corresponding deformation
of the free theory. For this aim we introduce three vectors
$|\chi^{(i)}\rangle_{s_i}$, gauge parameters
$|\Lambda^{(i)}\rangle_{s_i}$, $|\Lambda^{(i)1}\rangle_{s_i}$ with
corresponding vacuum vectors $|0\rangle^i$ and oscillators, where
$i=1,2,3$. Then one defines the deformed action and the deformed
gauge transformations in the form
\begin{eqnarray}\label{S[n]}
  && S_{[1]|(s)_3}[\chi^{(1)},\chi^{(2)}, \chi^{(3)}] \ = \  \sum_{i=1}^{3} \mathcal{S}_{0|s_i}   +
  g  \int \prod_{e=1}^{3} d\eta^{(e)}_0  \Big( {}_{s_{e}}\langle \chi^{(e)} K^{(e)}
  \big|  V^{(3)}\rangle_{(s)_{3}}+h.c. \Big)  , \\
   && \delta_{[1]} \big| \chi^{(i)} \rangle_{s_i}  =  Q^{(i)} \big| \Lambda^{(i)} \rangle_{s_i} -
g \int \prod_{e=1}^{2} d\eta^{(i+e)}_0  \Big( {}_{s_{i+1}}\langle
\Lambda^{({i+1})}K^{(i+1)}\big|{}_{s_{i+2}}
   \langle \chi^{({i+2})}K^{(i+1)}\big|  \label{cubgtr}\\
   &&
\ \   \phantom{\delta_{[1]} \big| \chi^{(i)} \rangle_{s_i}} +(i+1 \leftrightarrow i+2)\Big)
\big|\widetilde{V}{}^{(3)}\rangle_{(s)_{3}}, \nonumber\\
&& \delta_{[1]} \big| \Lambda^{(i)} \rangle_{s_i}  =  Q^{(i)} \big| \Lambda^{(i)1} \rangle_{s_i}
 -g  \int \prod_{e=1}^{2} d\eta^{(i+e)}_0  \Big( {}_{s_{i+1}}\langle \Lambda^{(i+1)1}K^{(i+1)}\big|{}_{s_{i+2}}
\langle \chi^{({i+2})}K^{(i+1)}\big|  \label{cubggtr}\\
   &&
\ \   \phantom{\delta_{[1]} \big| \chi^{(i)} \rangle_{s_i}} +(i+1
\leftrightarrow i+2)\Big) \big|\widehat{V}{}^{(3)}\rangle_{(s)_{3}}
\nonumber
\end{eqnarray}
with some yet unknown  vectors $\big| V^{(3)}\rangle_{(s)_{3}}, \,
\big|\widetilde{V}{}^{(3)}\rangle_{(s)_{3}}, \,
\big|\widehat{V}{}^{(3)}\rangle_{(s)_{3}}.$ Here
$\mathcal{S}_{0|s_i}$ is the free action (\ref{PhysStatetot}) for
the field $\big| \chi^{(i)} \rangle_{s_i}$,\, $Q^{(i)}$ is the BRST
charge corresponding to spin $s_{i},\, i=1,2,3$, $K^{(i)}$ is the
operator $K$ (\ref{geneq2}) corresponding to spin $s_{i},\, i=1,2,3$
and $g$ is a coupling constant. Also we use the notation $(s)_{3}
\equiv (s_1,s_2,s_3)$ and convention $[i+3 \simeq i]$.

A concrete construction of the cubic interaction means finding the
concrete vectors $\big| V^{(3)}\rangle_{(s)_{3}}$, $\big|
\widetilde{V}{}^{(3)}\rangle_{(s)_{3}}$,
$\big|\widehat{V}{}^{(3)}\rangle_{(s)_{3}}$. For this aim we can use
the set of fields, the constraints, ghost operators related with
spins $s_1,s_2,s_3$ and the condition of gauge invariance of the
deformed action under the deformed gauge transformations.

The integration over space-time coordinates is inherited from the
inner product definition (\ref{scalarprod}).  A local dependence on
space-time coordinates in the vertex $\big |V^{(3)}\rangle$, $\big |\widetilde{V}{}^{(3)}\rangle$ and
$\big |\widehat{V}{}^{(3)}\rangle$ implies
\begin{equation}\label{xdep}
  \big |V^{(3)}\rangle_{(s)_3} = \prod_{i=2}^3 \delta^{(d)}\big(x_{1} -  x_{i}\big) V^{(3)}_{(s)_3}
  \prod_{j=1}^3 \eta^{(j)}_0 |0\rangle , \ \  |0\rangle\equiv \otimes_{e=1}^3 |0\rangle^{e} .
\end{equation}
The conservation law for the momenta associated with vertices $\big
|V^{(3)}\rangle$, $\big|\widetilde{V}{}^{(3)}\rangle$ and
$\big |\widehat{V}{}^3\rangle$, holds true
\begin{equation}
\label{consermom}
 p^{(1)}_\mu +p^{(2)}_\mu +p^{(3)}_\mu  = 0.
\end{equation}
This relation will be used in the next section for explicit finding
the vertices.

Turn now to invariance  of the action $S_{[1]}$ with respect to
deformed transformations $\delta_{[1]} \big| \chi^{(i)}
\rangle_{s_i} $, $i=1,2,3.$   It yields to the following equations
in the zeroth and first orders in $g$ (the second order is not
required for finding the cubic vertex):
\begin{eqnarray}
g^0: && Q^{(i)}Q^{(i)}  |\Lambda^{(i)} \rangle_{s_i} =  \eta^{(i)+}_{11}\eta^{(i)}_{11}\sigma^{(i)} |\Lambda^{(i)} \rangle_{s_i}\equiv 0,   \ \ i=1,2,3,  \label{g0L} \\
g^1: &&   \int \prod_{e=1}^{3} d\eta^{(e)}_0   {}_{s_{j}}\langle
\Lambda^{(j)}K^{(j)} \big| {}_{s_{j+1}}\langle
\chi^{(j+1)}K^{(j+1)}\big|{}_{s_{j+2}}\langle
\chi^{(j+2)}K^{(j+2)}\big| \mathcal{Q}(V^3,\widetilde{V}^3) = 0,
\label{g1L}
\end{eqnarray}
where
\begin{eqnarray}
\phantom{g^1:} && \mathcal{Q}(V^3,\widetilde{V}^3) = \sum_{k=1}^3
Q^{(k)}\big |\widetilde{V}{}^{(3)}\rangle_{(s)_3}
   +  Q^{(j)}\Big( \big|{V}{}^{(3)}\rangle_{(s)_{3}} -\big|\widetilde{V}{}^{(3)}\rangle_{(s)_3}\Big)
    , \ j=1,2,3 .\label{g1operV3}   
\end{eqnarray}
The conservation of the form of the gauge transformations for the
fields $\big| \chi^{(i)} \rangle_{s_i} $ under the gauge
transformations $\delta_{[1]}\big| \Lambda^{(i)} \rangle_{s_i} $
with the parameters $\big| \Lambda^{(i)1} \rangle_{s_i} $ leads to
the similar equations
\begin{eqnarray}  g^0: && \hspace{-1em} Q^{(i)}Q^{(i)}  |\Lambda^{(i)1} \rangle_{s_i} =
\eta^{(i)+}_{11}\eta^{(i)}_{11}\sigma^{(i)} |\Lambda^{(i)1} \rangle_{s_i}\equiv 0,   \ \ i=1,2,3,
\label{g0L1} \\
g^1: && \hspace{-1em}  \int \prod_{e=1}^{2} d\eta^{(e)}_0   {}_{s_{j+1}}\langle \Lambda^{(j+1)1}K^{(j+1)}
\big| {}_{s_{j+2}}\langle \chi^{(j+2)}K^{(j+2)}\big|  \Big(\mathcal{Q}(\widetilde{V}^3,\widehat{V}^3)
- Q^{(j+2)} |\widehat{V}{}^{(3)}\rangle\Big) = 0,
\label{g1L1}
\end{eqnarray}
where
\begin{eqnarray}
\phantom{g^1:} && \mathcal{Q}(\widetilde{V}^3, \widehat{V}^3) =
\sum_{k=1}^3 Q^{(k)}\big |\widehat{V}{}^{(3)}\rangle_{(s)_3} +
Q^{(j)}\Big( \big|\widetilde{V}{}^{(3)}\rangle_{(s)_{3}}
-\big|\widehat{V}{}^{(3)}\rangle_{(s)_3} \Big) , \ j=1,2,3.
\label{g1operV32}   
\end{eqnarray}
Let us pay attention that the last term in the relation (\ref{g1L1})
can be omitted since it proportional to zero order in $g$ of right
hand side of the equation of motion $Q^{(j+2)}\big|
\chi^{(j+2)}\rangle$ and hence can be eliminated by field
redefinition in free action.

Taking into account the above deformed gauge transformations we,
should make sure that these transformations still form the closed
algebra, i.e. the following relation should be fulfilled
\begin{eqnarray}
&&  \big[\delta^{\Lambda_1}_{[1]},\delta^{\Lambda_2}_{[1]}\big]  |\chi^{(i)} \rangle \  =
\  - g \delta^{\Lambda_3}_{[1]}  |\chi^{(i)} \rangle  \ \
\label{closuregtr},
\end{eqnarray}
where the Grassmann-odd  gauge parameter  $\Lambda_3$ should be a
function of the parameters $\Lambda_1,\Lambda_2,$ $\Lambda_3 =
\Lambda_3(\Lambda_1,\Lambda_2).$ The validity of the relation
(\ref{closuregtr}) is verified by explicit calculations
\begin{eqnarray}&&  \big[\delta^{\Lambda_1}_{[1]},\delta^{\Lambda_2}_{[1]}\big]  |\chi^{(i)} \rangle
\  =  \ -   g  \bigg\{\Big(  \int \prod_{e=1}^{2} d\eta^{(i+e)}_0  \Big( \langle \Lambda^{(i+1)}_2K\big|\langle \Lambda^{(i+2)}_1\big|K Q^{(i+2)}  \label{commmgtr}   \\
&& \phantom{\big[\delta^{\Lambda_1}_{[1]},\delta^{\Lambda_2}_{[1]}\big]  |\chi^{(i)0} \rangle  \  =  \ } +\big(i+1 \leftrightarrow i+2\big)\Big)   - \big({\Lambda_1} \leftrightarrow {\Lambda_2}\big) \bigg\} \big|\widetilde{V}{}^{(3)}\rangle \nonumber \\
&& \phantom{\big[\delta^{\Lambda_1}_{[1]},\delta^{\Lambda_2}_{[1]}\big]  |\chi^{(i)0} \rangle} + g^2\bigg\{ \int \prod_{e=2}^{3} d\eta^{(i+e)}_0    \bigg[  \int \prod_{f=1}^{2}d\eta^{\prime(i+2+f)}_0 \langle \Lambda^{(i+1)}_2K\big|\Big(  \big|\Lambda^{(i+1)}_1\rangle \big|\chi^{(i)0}\rangle  \langle\widetilde{V}{}^{(3)} K^3\big| \nonumber \\
&&
\phantom{\big[\delta^{\Lambda_1}_{[1]},\delta^{\Lambda_2}_{[1]}\big]
|\chi^{(i)0} \rangle}  +\big(i \leftrightarrow
i+1\big)\Big)+\big(i+1 \leftrightarrow i+2\big)\bigg]  -
\big({\Lambda_1}  \leftrightarrow {\Lambda_2}\big) \bigg\}
\big|\widetilde{V}{}^{(3)}\rangle \nonumber
\end{eqnarray}
for $ K^3 \equiv  \otimes_{i=1}^3  K^{(i)}$. The relation
(\ref{commmgtr}) allows to find the parameter $\Lambda_3$ in cubic
approximation in the form
\begin{eqnarray}
\label{L3L1L2}
\hspace{-0.5em}&\hspace{-0.5em}& \hspace{-0.5em} \big|\Lambda_3
\rangle \sim  \int \prod_{e=1}^{2} d\eta^{(i+e)}_0  \Big( \langle
\Lambda^{(i+1)}_2K\big|\langle \Lambda^{(i+2)}_1\big|K  +\big(i+1
\leftrightarrow i+2\big)\Big)   - \big({\Lambda_1} \leftrightarrow
{\Lambda_2}\big) \bigg\} \big|\widetilde{V}{}^{(3)}\rangle .
\label{sollL3}
\end{eqnarray}
The condition of closure of the algebra of deformed gauge
transformations  provides the fact that a set of the generators of
deformed gauge transformations, encoded in (\ref{cubgtr}) will
remain complete, without appearance of new generators. In addition,
the commutator (\ref{closuregtr}) is the condition  which restricts
the form of the vertices $\big|\widetilde{V}{}^{(3)}\rangle$.

Also, the vertices should satisfy the spin conditions as the
consequence of the spin equation (\ref{extconstsp}) for each sample
$|\chi^{(i)}\rangle_{s_i},\,
 |\Lambda^{(i)}\rangle_{s_i},\, |\Lambda^{(i)1}\rangle_{s_i}$:
\begin{equation}\label{spinV}
 \sigma^{(i)}\Big(\big|{V}{}^{(3)}\rangle_{(s)_3},\, \big| \widetilde{V}{}^{(3)}\rangle_{(s)_3},\, \big| \widehat{V}{}^{(3)}\rangle_{(s)_3}\Big) = 0  ,
\end{equation}
providing the nilpotency of total BRST operator $Q^{tot} \equiv
\sum_i Q^{(i)}$ when evaluated on the vertices due to the equations
(\ref{g0L}) and $\{Q^{(i)}, Q^{(j)}\}=0$ for $i\ne j$. Indeed, from
the definition of the vertices $\big|{V}{}^{(3)}\rangle_{(s)_3}$
(\ref{S[n]}) it follows from the completeness of the inner product
that the  spin numbers $s_i$ for the oscillators  from each vector
${}_{s_{i}}\langle \chi^{(i)}\big|$ are equal to the spin numbers
$s_1, s_2, s_3$  of the oscillators in the each monomial from
$3$-vector  $\big|{V}{}^{(3)}\rangle_{(s)_3}$.

The equations (\ref{spinV}) and (\ref{g1operV3}), (\ref{g1operV32}):
\begin{equation}\label{g1oper}
\mathcal{Q}(V^3,\widetilde{V}^3)  = 0, \qquad
\mathcal{Q}(\widetilde{V}^3, \widehat{V}^3) =0,
\end{equation}
together  with the form of the commutator of the gauge
transformations (\ref{commmgtr}) determine the cubic  interacting
vertices for irreducible  massless totally symmetric higher spin
fields.

\section{General solution for cubic vertices}
\label{intBRSTBFV}

In this section we will construct the general solution for the cubic
vertex.

Further, for simplicity, we assume that
$\big|{\widetilde{V}}{}^{(3)}\rangle_{(s)_3}$ =
$\big|{V}{}^{(3)}\rangle_{(s)_3}$ =
$\big|\widehat{V}{}^{(3)}\rangle_{(s)_3}.$ Then\footnote{In
principle, the equations (\ref{g1L}), (\ref{g1L1}) can be solved
without these simplifying assumption however, it complicates the
consideration.}, the equations (\ref{g1L}), (\ref{g1L1}) and the
operators (\ref{g1operV3}), (\ref{g1operV32}) take the form
\begin{equation}
\label{g1Lmod}
\left\{ \begin{array}{c}
{}_{s_{2}}\langle \Lambda^{(1)}K^{(1)}\big| {}_{s_{2}}\langle \chi^{(2)}K^{(2)}\big|{}_{s_3}\langle
\chi^{(3)}K^{(3)}\big| \mathcal{Q}({V}^3,{V}^3)  =0 \\
{}_{s_{j+1}}\langle \Lambda^{(j+1)1}K^{(j+1)}\big|
{}_{s_{j+2}}\langle \chi^{(j+2)}K^{(j+1)}\big|
\mathcal{Q}({V}^3,{V}^3) = 0
\end{array} \right.   \Longrightarrow \  Q^{tot}
\big|{V}{}^{(3)}\rangle_{(s)_3}  =0.
\end{equation}
Remind that $Q^{tot} \equiv \sum_i Q^{(i)}.$

We look for a general solution of the equation (\ref{g1Lmod}) in the
form of products of specific operators, homogenous in oscillators.
First, they include the different $Q^{tot}$- BRST closed  forms
$\mathcal{L}^{(i)}_{k_i}$, $i=1,2,3$, $k_i = 1, ..., s_i$ and $\mathcal{Z}$ constructed from ones ${L}^{(i)}$, $Z$ known from \cite{BRST-BV3}, where
$L^{(i)}$ are the linear in oscillators and $Z$ is cubic in
oscillators,
\begin{eqnarray}\label{LrZLL}
  &&  \mathcal{L}^{(i)}_{k_i} \ = \  ({L}^{(i)})^{k_i-2} \Big(({L}^{(i)})^{2} -  \frac{\imath k_i!}{2(k_i-2)!} \eta_{11}^{(i)+} \big[ 2\mathcal{P}^{(i+1)}_0+ 2 \mathcal{P}^{(i+2)}_0-\mathcal{P}^{(i)}_0
\big] \Big)  ,  \\
    &&  L^{(i)} \ = \   (p^{(i+1)}_{\mu}-p^{(i+2)}_{\mu})a^{(i)\mu+} - \imath \big(\mathcal{P}^{(i+1)}_0- \mathcal{P}^{(i+2)}_0
  \big)\eta_1^{(i)+} , \nonumber \\
  && Z \ = \  L^{(12)+}_{11}L^{(3)} + L^{(23)+}_{11}L^{(1)} + L^{(31)+}_{11}L^{(2)}. \label{LrZ1}
\end{eqnarray}
Here we have used the relations (\ref{Lr11})--(\ref{Lr11Lr2}) below,  $p^{(i)}_{\mu} = -i\partial^{(i)}_{\mu}$ and
\begin{eqnarray}
L^{(i i+1)+}_{11} \ = \ \textstyle a^{(i)\mu+}a^{(i+1)+}_{\mu} -
\frac{1}{2}\mathcal{P}^{(i)+}_1\eta_1^{(i+1)+} -
\frac{1}{2}\mathcal{P}^{(i+1)+}_1\eta_1^{(i)+}.\label{Lrr+1}
 \end{eqnarray}
Second, these operators involve the  new two-, four- , ..., $[s_i
/2]$ forms in powers of oscillators, corresponding to the trace
operators
\begin{equation}\label{trform}
U^{(s_i)}_{j_i}\big(\eta_{11}^{(i)+},\mathcal{P}_{11}^{(i)+} \big) \
: = \
(\widehat{L}{}^{(i)+}_{11})^{(j_i-2)}\big\{(\widehat{L}{}^{+(i)}_{11})^2
- j_i(j_i-1)\eta_{11}^{(i)+}\mathcal{P}_{11}^{(i)+}\big\}, \ \
i=1,2,3.
\end{equation}
First of all we have to check that the operators (\ref{LrZLL}) and
(\ref{Lrr+1}) are closed relatively operator $Q^{tot}$ which is
extension of the BRST operator in works \cite{reviews3},
\cite{BRST-BV3} by the operators responsible for the traces. Indeed,
the $Q^{tot}$- BRST closeness for the operator $L^{(i)}$ is reduced
to the fulfillment of the equations at the terms linear in
$\eta_{11}^{(i)+}$
 \begin{eqnarray}\label{Lr11}
&&       \widehat{L}{}^{(i)}_{11}(L^{(i+j)})^k |0\rangle  \equiv 0, j=1,2,  \  \  \forall k \in \mathbb{N}\,,  \\
&&     \widehat{L}{}^{(i)}_{11}L^{(i)} |0\rangle =  \Big(- (p^{(i+1)}_{\mu}-p^{(i+2)}_{\mu})a^{(i)\mu}+ \imath \big(\mathcal{P}^{(i+1)}_0- \mathcal{P}^{(i+2)}_0\big)\eta_1^{(i)} + L^{(i)}\widehat{L}{}^{(i)}_{11}\Big)|0\rangle =0 , \label{Lr11Lr1} \\
&&    \widehat{L}{}^{(i)}_{11}(L^{(i)})^2 |0\rangle = \Big(- (p^{(i+1)}_{\mu}-p^{(i+2)}_{\mu})a^{(i)\mu}+ \imath \big(\mathcal{P}^{(i+1)}_0- \mathcal{P}^{(i+2)}_0\big)\eta_1^{(i)}+ L^{(i)}\widehat{L}{}^{(i)}_{11}]\Big)L^{(i)}|0\rangle , \label{Lr11Lr2}\\
&&\phantom{L^{(r)}_{11}} =  \Big( \eta^{\mu\nu}(p^{(i+1)}_{\mu}-p^{(i+2)}_{\mu})  (p^{(i+1)}_{\nu}-p^{(i+2)}_{\nu}) + L^{(i)}\imath \big(\mathcal{P}^{(i+1)}_0- \mathcal{P}^{(i+2)}_0\big)\eta_1^{(i)}+ (L^{(i)})^2\widehat{L}{}^{(i)}_{11}]\Big)|0\rangle\nonumber \\
&& \phantom{L^{(r)}_{11}} =    \eta^{\mu\nu}(p^{(i+1)}_{\mu}-p^{(i+2)}_{\mu})  (p^{(i+1)}_{\nu}-p^{(i+2)}_{\nu})|0\rangle \ne 0.\nonumber
 \end{eqnarray}
The last relations and ones for $(L^{(i)})^k $ do not vanish under the sign of inner
products and justify the introduction of the forms (\ref{LrZLL}). By the same reason, the  any power  of the
form $Z$ is not BRST-closed as well.

Let us prove now BRST-closeness of the new forms (\ref{trform}). To
see that it is sufficient to present the vertex as
$\big|V^{(3)}\rangle_{(s)_3} $ $=
U^{(s_i)}_{j_i}\big|X^{(3)}\rangle_{(s)_3-2j_i} $ with some
$\big|X^{(3)}\rangle_{(s)_3-2j_i}$, where
 $(\varepsilon,
gh)\big|X^{(3)}\rangle$ = $(1,3)$ and check the closeness
for $i\ne k$ and  for $j=1,2$
 \begin{eqnarray}\label{Lr11L+}
&&     \eta_{11}^{(i)+}    \widehat{L}{}^{(i)}_{11} U^{(s_k)}_{j_k} |0\rangle  \equiv 0,  k\ne i \,,  \\
&&     \eta_{11}^{(i)+} \widehat{L}{}^{(i)}_{11} U^{(s_i)}_{1_i}\big|X^{(3)}\rangle_{(s)_3-2_i} = \big(\sigma^{(i)} \eta_{11}^{(i)+}  - 2\eta_{11}^{(i)+}\mathcal{P}_{11}^{(i)+}\eta_{11}^{(i)}\big)\big|X^{(3)}\rangle_{(s)_3-2_i} = 0,  \label{Lr11Lr11} \\
&&   \eta_{11}^{(i)+}  \widehat{L}{}^{(i)}_{11} U^{(s_i)}_{2_i}\big|X^{(3)}\rangle_{(s)_3-4_i} =   \big(\{\sigma^{(i)} \eta_{11}^{(i)+} , \widehat{L}{}^{(i)+}_{11}\} -2\eta_{11}^{(i)+} \widehat{L}{}^{(i)+}_{11} + 2\eta_{11}^{(i)+} \widehat{L}{}^{(i)+}_{11}   \label{Lr11U2}\\
&& \phantom{\eta_{11}^{(i)+}  L^{(i)}_{11}} -
4\widehat{L}{}^{(i)+}_{11}\eta_{11}^{(i)+}
\mathcal{P}_{11}^{(i)+}\eta_{11}^{(i)}
\big)\big|X^{(3)}\rangle_{(s)_3-4_i} \  = \ 2\sigma^{(i)}
\eta_{11}^{(i)+}  \widehat{L}{}^{(i)+}_{11}
\big|X^{(3)}\rangle_{(s)_3-4_i} =  0. \nonumber
\end{eqnarray}
For arbitrary $j>2$ we have
 \begin{eqnarray}
&&  \eta_{11}^{(i)+} \widehat{L}{}^{(i)}_{11}U^{(s_i)}_{j_i}\big|X^{(3)}\rangle_{(s)_3-2j_i} =   \Big(\sum_{e=0}^{jj_i-1}(\widehat{L}{}^{(i)+}_{11})^{e}\{\sigma^{(i)} \eta_{11}^{(i)+} , \widehat{L}^{(i)+}_{11}\}(\widehat{L}^{(i)+}_{11})^{j_i-e-2} \label{Lr11Uj} \\
&& \phantom{\eta_{11}^{(i)+}  L^{(i)}_{11}U^{(s_i)}_{2_i}}   + j_i\eta_{11}^{(i)+}\big[(j_i-1)   -  2\mathcal{P}_{11}^{(i)+}\eta_{11}^{(i)}\big](\widehat{L}{}^{(i)+}_{11})^{j_i-1} \Big) \big|X^{(3)}\rangle_{(s)_3-2j_i} \nonumber \\
&&  \phantom{\eta_{11}^{(i)+}  L^{(i)}_{11}U^{(s_i)}_{2_i}}  =  \Big( (j_i-1)  \sigma^{(i)}\eta_{11}^{(i)+} - 2j_i\eta_{11}^{(i)+}\mathcal{P}_{11}^{(i)+}\eta_{11}^{(i)}\Big) \big(\widehat{L}^{(i)+}_{11}\big)^{j_i-1} \big|X^{(3)}\rangle_{(s)_3-2j_i} =      0 . \nonumber
 \end{eqnarray}
In the relations (\ref{Lr11L+})--(\ref{Lr11Uj}) we have used the
supercommutators which linear in $\widehat{L}^{(i)+}_{11}$
\begin{eqnarray}\label{auxrel}
&& \eta_{11}^{(i)+}  \left[ \widehat{L}{}^{(i)}_{11} , \, \widehat{L}^{(i)+}_{11} \right\} = \eta_{11}^{(i)+} \big\{ (\sigma^{(i)} + 2) - 2\mathcal{P}_{11}^{(i)+}\eta_{11}^{(i)}\big\} =  \sigma^{(i)} \eta_{11}^{(i)+} - 2\eta_{11}^{(i)+}\mathcal{P}_{11}^{(i)+}\eta_{11}^{(i)},\\
&&   \left[ \widehat{L}{}^{(i)+}_{11} , \, \sigma^{(j)}\right\} = - 2\delta^{ij} \widehat{L}{}^{(i)+}_{11} , \label{auxrel0}
\end{eqnarray}
then, polynomial in $\widehat{L}^{(i)+}_{11}$
\begin{eqnarray}
&& \eta_{11}^{(i)+}  \left[ \widehat{L}{}^{(i)}_{11} , \, \big(\widehat{L}^{(i)+}_{11}\big)^{j_i} \right\} =   \sum_{e=0}^{j_i-1}\big(\widehat{L}^{(i)+}_{11}\big)^e\eta_{11}^{(i)+}  \left[ \widehat{L}{}^{(i)}_{11} , \, \widehat{L}^{(i)+}_{11} \right\} \big(\widehat{L}^{(i)+}_{11}\big)^{j_i-e-1} \label{auxrel1}\\
&& \phantom{\eta_{11}^{(i)+}  \left[ \widehat{L}{}^{(i)}_{11}  \right\}} = \sum_{e=0}^{j_i-1}\big(\widehat{L}^{(i)+}_{11}\big)^e  \sigma^{(i)} \eta_{11}^{(i)+} \big(\widehat{L}^{(i)+}_{11}\big)^{j_i-e-1}- 2j_i\big(\widehat{L}^{(i)+}_{11}\big)^{j_i-1}\eta_{11}^{(i)+}\mathcal{P}_{11}^{(i)+}\eta_{11}^{(i)}.\nonumber \\
&& \phantom{\eta_{11}^{(i)+}  \left[ \widehat{L}{}^{(i)}_{11}  \right\}}
=(j_i-1)\Big[  \sigma^{(i)} -j_i\Big]\eta_{11}^{(i)+}\big(\widehat{L}^{(i)+}_{11}\big)^{j_i-1} -
 2j_i\big(\widehat{L}^{(i)+}_{11}\big)^{j_i-1}\eta_{11}^{(i)+}\mathcal{P}_{11}^{(i)+}
 \eta_{11}^{(i)}\nonumber .
\end{eqnarray}
Besides we took into account an independence of
$\big|X^{(3)}\rangle_{(s)_3-2j_i}$ on the momenta
$\mathcal{P}_{11}^{(i)+}$ and the fact that
\begin{equation}\label{obv}
\sigma^{(i)}
\eta_{11}^{(i)+}\big(\widehat{L}^{(i)+}_{11}\big)^{j_i-1}\big|X^{(3)}\rangle_{(s)_3-2j_i}
=  0.
\end{equation}

As the result, the general solution for the equation (\ref{g1Lmod})
for cubic vertex has the form
\begin{eqnarray}\label{genvertex}
   |{V}{}^{(3)}\rangle_{(s)_3} &=& |{V}{}^{M(3)}\rangle_{(s)_3}  + \sum_{(j_1,j_2,j_3) >0}^{([s_{1}/2],[s_{2}/2],[s_{3}/2])} U^{(s_1)}_{j_1}U^{(s_2)}_{j_2}U^{(s_3)}_{j_3}|{V}{}^{M(3)}\rangle_{(s)_3-2(j)_3},
   \end{eqnarray}
where the vertex $\big|{V}{}^{M(3)}\rangle_{(s)_3-2(j)_3}$ was
defined in Metsaev's paper \cite{BRST-BV3} with account for (\ref{xdep})  but with modified forms $\mathcal{L}^{(i)}_{k_i}$,  (\ref{LrZLL})  and $\mathcal{Z}_j$ instead of ${Z}^j$ (\ref{LrZ1})
\begin{eqnarray}\label{Vmets}
{V}{}^{M(3)}_{{(s)_3-2(j)_3}} & = &   \sum_{k}\mathcal{Z}_{1/2\{(s-2J) - k\}}\prod_{i=1}^3 \mathcal{L}^{(i)}_{s_{i}-2j_i-1/2(s-2J- k)} ,  \\
(s,J) &= & \big(\sum_{i}s_i , \ \sum_ij_i\big).
\end{eqnarray}
and is numerated by the natural parameter $k$ subject to the
equations
 \begin{eqnarray}
  s-2J-2s_{\min}\leq k\leq s-2J, \ \ \ k=s-2J - 2p, \ p\in \mathbb{N}_0.
\end{eqnarray}
Here the quantity $\mathcal{Z}_j$ is determined, e.g.  for $j=1$ as follows
\begin{eqnarray}
  && \mathcal{Z}\prod_{j=1}^3\mathcal{L}^{(j)}_{k_j} = {Z}\prod_{j=1}^3\mathcal{L}^{(j)}_{k_j} - \sum_{i=1}^3k_i\frac{b^{(i)+}}{h^{(i)}}\big[\big[\widehat{L}^{(i)}_{11},{Z}\big\},{L}^{(i)} \big\}\prod_{j=1}^3\mathcal{L}^{(j)}_{k_j-\delta_{ij}} \nonumber \\
&& +\sum_{i\ne e}^3 k_ik_e\frac{b^{(i)+}b^{(e)+}}{h^{(i)}h^{(e)}}\big[\widehat{L}^{(e)}_{11},\big[\big[\widehat{L}^{(i)}_{11},{Z}\big\},{L}^{(i)} \big\}{L}^{(e)} \big\}\prod_{j=1}^3\mathcal{L}^{(j)}_{k_j-\delta_{ij}-\delta_{ej}} \nonumber\\
&& - \sum_{i\ne e\ne o}^3k_ik_ek_o\frac{b^{(i)+}b^{(e)+}b^{(o)+}}{h^{(i)}h^{(e)}h^{(o)}}\big[\widehat{L}^{(o)}_{11},\big[\widehat{L}^{(e)}_{11},\big[\big[\widehat{L}^{(i)}_{11},{Z}\big\},{L}^{(i)} \big\} {L}^{(e)} \big\}{L}^{(o)} \big\}\prod_{j=1}^3\mathcal{L}^{(j)}_{k_j-1}. \nonumber
\end{eqnarray}
For $j>1 $ the expressions for $\mathcal{Z}_j$ may be derived in the similar way.

The general vertex (\ref{genvertex}) contains besides the known modified term
(\ref{Vmets}) the new terms. These new vertices have the following
structure. First, they contain the linear terms in powers of trace
operators $U^{(s_i)}_{1_i}= \widehat{L}{}^{(i)+}_{11}$  at least for
one higher spin field copy
\begin{eqnarray}
   \sum_{(j_1,j_2,j_3) > 0}^{(1,1,1)}  (\widehat{L}{}^{(1)+}_{11})^{j_1}(\widehat{L}{}^{(2)+}_{11})^{j_2}(\widehat{L}{}^{(3)+}_{11})^{j_3}|{V}{}^{M(3)}\rangle_{(s)_3-2(j)_3}.  \end{eqnarray}
Note, that in comparison with cubic vertex constructed in
\cite{BRST-BV3}, the general cubic vertex contains the degree 2
homogeneous polynomials in the oscillators
$\widehat{L}{}^{+(1)}_{11}$, $\widehat{L}{}^{+(2)}_{11}$,
$\widehat{L}{}^{+(3)}_{11}$, which depend  in addition on $b^{(1)+}$, $b^{(2)+}$,
$b^{(3)+}$,
oscillators respectively.  Second, the quantity $U^{(s_i)}_{j_i}$ (\ref{trform})
for $j_i\geq 2$ depends on the product
$\eta_{11}^{(i)+}\mathcal{P}_{11}^{(i)+}$ with spin value $4$.

Therefore, for even $s_i$, $i=1,2,3$  the  vertex for $ j_i =
[s_{i}/2]$ will contain the $s_{i}/2$ traces for the initial fields
$|\phi\rangle_{s_{i}}$, $\prod_{i} Tr^{s_{i}/2}\phi^{(i)}$,  without
any derivatives
\begin{equation}\label{wder}
\overline{V}^{(3)}_{(s)_3}\  =  \  \prod_{i=1}^3U^{(s_i)}_{[s_{i}/2]} =  \prod_{i=1}^3(\widehat{L}{}^{(i)+}_{11})^{(j_i-2)}\big\{(\widehat{L}{}^{+(i)}_{11})^2  - j_i(j_i-1)\eta_{11}^{(i)+}\mathcal{P}_{11}^{(i)+}\big\} .
\end{equation}
In case of one, two or all odd values of the helicities
$s_1,s_2,s_3,$ the vertices with  minimal number of derivatives will
contain respectively  one, two or three derivatives:
\begin{eqnarray}\label{minder1}
&&  {V}^{(3)}_{1{}(s)_3}\  =  \  U^{(s_1)}_{[s_1-1/2]}U^{(s_2)}_{[s_2/2]}U^{(s_3)}_{[s_3/2]}  L^{(1)}  ,\\
\label{minder2}
&&  {V}^{(3)}_{2{}(s)_3}\  =  \  U^{(s_1)}_{[s_1-1/2]}U^{(s_2)}_{[s_2-1/2]}U^{(s_3)}_{[s_3/2]}  L^{(1)}L^{(2)}  ,\\
\label{minder3}
&&  {V}^{(3)}_{3{}(s)_3}\  =  \  \prod_{i=1}^3U^{(s_i)}_{[s_i-1/2]} \big\{\prod_{i=1}^3 L^{(i)}+ Z \big\}.
\end{eqnarray}
For $d>4$ the number of independent (parity invariant)  vertices in
each $ |{V}{}^{M(3)}\rangle_{{(s)_3-2(j)_3}}$ enumerated  by $k$  is
equal to $(s_{\min}+1)$, whereas for $d=4$ it reduced to  $2$ , i.e.
$k=s-2J,\, s-2J-2s_{min} $ due to proportionality of the vertices
for $k$: $s-2J-2s_{min} <k<s-2J,$ to the terms with d'Alambert
operator $\partial^2$ which can be removed by the field
redefinitions (see \cite{Metsaev0512} for the explanation). For
equal helicities, $s_1=s_2=s_3$ there is the cubic vertex
(\ref{genvertex}) for self-interacting higher spin fields
$\phi_{\mu(s)}$.

Emphasize, that the including the constraints $L_{11}$ responsible
for the traces into BRST operator means that the standard condition
of vanishing double traces of the fields is fulfilled only on-shell
as the consequence of free equations of motion. Off-shell the
(double) traces of the fields do not vanish. As the result, the
trace constraints come into the cubic vertices. To pass from our
completely irreducible formulation to the formulation, where the
trace constraints are not included into BRST charge, we should use
the free equations of motion.

\section{Conclusion}

\label{Discus} 

To summarize, we have constructed generic cubic vertices for an
irreducible gauge-invariant Lagrangian formulation of massless
totally-symmetric higher spin  fields with arbitrary helicities
$s_1$, $s_2$ $s_3$ in $d$-dimensional  Minkowski space-time. The
construction is realized within the BRST approach to higher spin
field theories, and, unlike all the previous works, every constraint
that determines an irreducible massless higher spin representation
has been taken into account on equal footing in a complete BRST
operator.

To find cubic vertices being consistent with a deformed gauge
invariance, we have followed an additive deformation of classical
actions for three copies of the higher spin fields and the gauge
transformations for the fields and gauge parameters, while demanding
for the deformed action  to be invariant in a linear approximation
with respect to the coupling constant $g$, and for the gauge algebra
to be closed on a deformed mass shell up to the second order in $g$.
These requirements lead to a system of generating equations for the
cubic vertices, containing the total BRST invariance operator
condition $\mathcal{Q}(V^3,\widetilde{V}^3) =0$,
$\mathcal{Q}(\widetilde{V}^3,\widehat{V}{}^3) =0$ (\ref{g1operV3}),
(\ref{g1operV32}), the spin condition (\ref{spinV}), and the
condition (\ref{closuregtr}) for the gauge algebra closure. The
cubic vertex, in the particular case of coinciding vertices,
${V}^3=\widetilde{V}^3 =\widehat{V}{}^3$, satisfies the  equations
(\ref{spinV}), (\ref{g1Lmod}), and their solution is found  using a
set of BRST-closed forms, which include the modified forms (\ref{LrZLL}),
(\ref{LrZ1}),  constructed from ones in \cite{BRST-BV3}, and the new forms
(\ref{trform}) related to the trace operator constraints, having
dependence on additional oscillators, $b^{(i)+}$,
$\eta_{11}^{(i)+}$, $\mathcal{P}_{11}^{(i)+}$. As a result,
admissible cubic vertices for irreducible fields may involve terms
with less space-time derivatives in comparison with \cite{BRST-BV3}.

A general  solution to the equation (\ref{g1Lmod}) for a cubic
vertex  is presented by (\ref{genvertex}) and contains, in addition
to the familiar vertices (containing derivatives) given in
\cite{BRST-BV3}, some new terms containing traces in the BRST-closed
operators $U^{(s_i)}_{j_i}$ of the rank $j=1,2,..., [s_i/2]$
without derivatives. For even helicities $s_i$, $i=1,2,3,$, the
vertex for $ j_i =  [s_{i}/2]$ in (\ref{wder}) thereby contains
$s_{i}/2$ traces for the initial fields $|\phi\rangle_{s_{i}}$
without any derivatives. In  the case of one, two, or all odd values
of helicities, $s_1,s_2,s_3,$, the vertices with a minimal
number of derivatives will contain one
(\ref{minder1}), two (\ref{minder2}) or three (\ref{minder3})
derivatives, respectively. The results of \cite{BRST-BV3}
can be obtained from our results if one switches off all the terms
in the cubic vertex (\ref{genvertex}) that contain the operators
$\widehat{L}_{11}^{(i)+}$ and $ \eta_{11}^{(i)+}\mathcal{P}_{11}^{(i)+}$,
responsible for the trace constraints.

There are numerous directions for application and development  of
the suggested approach, such as finding cubic vertices for
irreducible massless half-integer higher spin fields on flat
backgrounds; for massive integer and half integer higher spin
fields, for higher spin fields with mixed symmetry of indices, for
higher spin supersymmetric fields, where the vertices should include
any powers of traces. The construction under consideration can be
generalized to find cubic vertices for irreducible higher spin
fields on anti-de-Sitter backgrounds, having in mind the
impossibility of making a flat limit for many of  the cubic vertices
in anti-de-Sitter spaces \cite{FradkinVasiliev},
\cite{FradkinVasiliev1}. One should also note the problems of
constructing the fourth and higher vertices within the BRST
approach. The interesting results in this direction were obtained in
\cite{DT}. We plan to address all of the mentioned problems in our
forthcoming works.

\vspace{-1ex}

\paragraph{Acknowledgements}
The authors are grateful to R.R. Metsaev and V.A. Krykhtin    for useful comments. The
work was partially supported by the Ministry of Education of Russian
Federation, project No FEWF-2020-003.

\end{document}